# High-Q Magneto-Optical Microdisk Resonators Based on Monocrystalline Bismuth-substituted Yttrium Iron Garnets

**Authors:** [1,5]Takeru Yambe, [1]Kota Taniguchi, [1]Tatsuya Kitai, [1]Daisuke Sato, [1]Syuan Gao, [2,3]Hajime Kumazaki, [2]Shun Fujii, [3]Takasumi Tanabe, [4]Satoshi Iwamoto, [1]Yasutomo Ota*.

[1]Department of Applied Physics and Physico-Informatics, Keio University, Yokohama Kanagawa 223-8522, Japan.

[2]Department of Physics, Faculty of Science and Technology, Keio University, Yokohama Kanagawa 223-8522, Japan.

[3]Department of Electronics and Electrical Engineering, Faculty of Science and Technology, Keio University, Yokohama Kanagawa 223-8522, Japan.

[4]Research Center for Advanced Science and Technology and Industrial Institute of Science, University of Tokyo, Tokyo 1538904, Japan

[5]Takeru Yambe. Email: takeru.yambe@keio.jp

*Yasutomo Ota. Email: ota@appi.keio.ac.jp

**Abstract:** Microresonators based on diverse material platforms have driven the progress of integrated photonics by enhancing light-matter interactions in compact volumes. Extending this strategy to magneto-optical (MO) media is highly attractive for nonreciprocal photonics, magnonics, and quantum transduction, yet compact high$Q$ resonators based on single-crystalline yttrium iron garnet (YIG) have remained elusive because of the difficulty of its nanofabrication. Here we demonstrate high-$Q$, small-mode-volume MO microdisk resonators based on monocrystalline Bi-substituted YIG (Bi:YIG). Low-loss Bi:YIG thin films prepared by bonding and thinning, combined with optimized Ar-plasma etching, enable submicron-thick, air-suspended microdisks that support telecomband whispering gallery modes with $Q = 1.52 \times 10^5$, a mode volume $V \approx 200\ (\lambda/n)^3$, and $Q/V \approx 760$, representing an orders-of-magnitude improvement for integrated high-$Q$ YIG MO cavities. The device exhibits clear magnetic-field-dependent spectral responses, including reciprocal resonance shifts and nonreciprocal frequency splitting of counter-propagating modes. A perturbative analytical model quantitatively reproduces these responses and indicates an effective MO coefficient about twice the bulk value, suggesting a possible miniaturization-induced enhancement. These results establish monocrystalline YIG microdisks as a compact high-$Q$ MO platform for integrated isolators, optomechanics, microcombs, and photon-magnon quantum interfaces.

## 1. Introduction

Optical microresonators have been a central engine of integrated photonics because they store light for long photon lifetimes in small mode volumes ($V$), thereby amplifying otherwise weak light–matter interactions[1]. Among various cavity geometries, microdisk resonators (MDRs) are especially attractive: their simple planar structure supports high-$Q$ whispering-gallery modes (WGMs), lithographic scalability, and efficient coupling to integrated waveguides or tapered fibers. The continual expansion of MDRs into new material systems has directly broadened the functionality of integrated photonics. Silica, silicon nitride, lithium niobate, compound semiconductors, and other material platforms have enabled optical frequency combs[2–4], nonlinear wavelength conversion[5–7], low-threshold lasers[8–10], quantum photonic interfaces[11,12], and precision sensing[13]. In this context, developing high-$Q/V$ MDRs from unexplored functional materials is not only a fabrication challenge, but also a route to new photonic physics and devices.

Magneto-optical (MO) materials are a particularly important target for this strategy. Their broken time-reversal symmetry enables nonreciprocal optical responses, which are essential for optical isolators and circulators, while their magnetic degrees of freedom provide access to magnon-mediated optical functionalities. Single-crystalline yttrium iron garnet (YIG) is one of the most promising MO media because it combines optical transparency at telecommunication wavelengths, appreciable room-temperature MO effects[14], and excellent spin coherence[15]. If the intrinsic material quality of single-crystalline YIG could be combined with the strong optical confinement of an integrated MDR, the resulting high-$Q/V$ platform would be highly beneficial for nonreciprocal integrated photonics[16–19], dark soliton microcombs[20,21], cavity optomagnonics[22–28], and microwave-to-optical quantum transducers[29].

Realizing such a platform, however, has remained difficult. Conventional epitaxial YIG films on gadolinium gallium garnet (GGG) provide high crystalline quality but weak vertical optical confinement because of their low refractive index contrast, whereas deposited or deeply patterned garnet films often suffer from degraded crystallinity or fabrication-induced optical loss[30]. Previous YIG-based micro/nanophotonic structures, including magnetoplasmonic devices[31–33], all-dielectric metasurfaces[34–36], photonic crystal cavities[37,38], and Fabry–Perot-type resonators[39,40], have therefore faced a trade-off between optical confinement, $Q$ factor, and material quality. In contrast, macroscopic YIG spheres can support high-$Q$ WGMs and have played a major role in cavity optomagnonics, but their large mode volumes and non-planar geometry limit their suitability for dense on-chip integration[22,24,29]. A planar, submicron-thick, single-crystalline YIG microdisk would overcome these limitations by retaining the favorable material properties while reducing $V$ by orders of magnitude.

Here, we demonstrate high-$Q$, small-$V$ MO microdisk resonators based on monocrystalline Bi-substituted YIG (Bi:YIG). By preparing low-loss Bi:YIG thin films through bonding and thinning, and patterning them with optimized Ar-plasma etching, we realize submicron-thick air-suspended MDRs that support telecom-band WGMs. The fabricated devices exhibit cavity resonances around 1550 nm with a maximum $Q$ of $1.52 \times 10^5$, together with a $V \approx 200\ (\lambda/n)^3$ and $Q/V \approx 720$, representing an orders-of-magnitude improvement for integrated high-$Q$ YIG MO cavities. We further observe clear magnetic-field-dependent spectral responses, including reciprocal resonance shifts and nonreciprocal frequency splitting between counter-propagating modes. These responses are quantitatively reproduced by a perturbative analytical model, which suggests an effective MO coefficient about twice the bulk value. These results establish monocrystalline YIG MDRs as a compact high-$Q/V$ MO cavity platform that links integrated photonics, nonreciprocal optics, and magnon-based light–matter interactions.

## 2. Device Fabrication

Fabricating high-$Q/V$ nanophotonic devices from monocrystalline YIG requires overcoming two coupled challenges: preparing a thin single-crystalline membrane that provides strong optical confinement, and patterning it without introducing excessive scattering loss. Conventional YIG films on GGG are limited by the small refractive-index contrast and chemical similarity between the two materials, which make tight vertical light confinement and membrane isolation difficult[35]. Alternative deposition-based approaches improve process flexibility but often compromise crystallinity[41,42]. In addition, the mechanical hardness and chemical inertness of YIG make low-damage micropatterning challenging; previous approaches frequently suffered from redeposition, rough sidewalls, or surface damage[38,43–45]. To address these issues, we have recently developed a nanofabrication approach for monocrystalline YIG nanophotonic[46]. The process consists of two key steps: the formation of a YIG-on-insulator (YIGOI) platform through bonding and thinning, and subsequent high-precision micropatterning by Ar-plasma ion etching. Using this approach, we prepared submicron-thick monocrystalline Bi:YIG membranes and patterned them into air-suspended MDRs while suppressing etch-induced sidewall roughness.

Figure 1(a) shows the fabrication process for the YIGOI substrate. We used a 225-μm-thick Bi-substituted YIG wafer grown by liquid-phase epitaxy (GLB, GRANOPT). Bi substitution enhances the MO response of the garnet material[47]. The Bi:YIG wafer was first cleaned with isopropanol and bonded to a thermally oxidized Si substrate consisting of a 2-μm-thick $SiO_2$ layer on a 625-μm-thick Si wafer. After bonding, the sample was annealed to strengthen the adhesion through a dehydration reaction. The Bi:YIG layer was then mechanically polished down to approximately 10 μm, followed by chemical mechanical polishing (CMP) to reduce surface roughness. Finally, the film thickness was further reduced to 150–330 nm by Ar-based plasma etching. The etching rate of approximately 1.1 nm/s allowed thickness control with a precision of roughly 10 nm. Figure 1(c) presents an optical image of a completed YIGOI substrate and its thickness distribution measured by white-light interferometric profilometry (Optical NanoGauge C10027-02, Hamamatsu Photonics Corporation). Macroscopic color fringes are visible across the film, reflecting thickness variations introduced during mechanical polishing. Nevertheless, the central region reached the target thickness of approximately 300 nm. The local thickness variation near this region was gradual, with a gradient of only 0.03 nm/μm. This small variation over the device scale has a negligible impact on the patterning process and optical resonator performance.

Figure 1(b) shows the process flow for fabricating air-suspended Bi:YIG MDRs. The disk patterns were drawn by electron-beam lithography using a positive resist (ZEP520A, ZEON Corporation). A conductive polymer layer (Espacer 300Z, Resonac) was coated on the resist to suppress charging during electron-beam exposure. The resist pattern was then transferred into the Bi:YIG layer by Ar-based inductively coupled plasma reactive ion etching (ICP-RIE) with an ICP power of 500 W, a bias power of 200 W, and a chamber pressure of 3 Pa. In designing the lithographic pattern, we intentionally minimized the exposed etching area around each disk to suppress redeposition of sputtered YIG material, which is a major source of sidewall roughness in Ar-based dry etching. Experiments using bulk YIG substrates with different opening widths confirmed that reducing the exposed etching area effectively suppresses redeposition around the disk structures (see Supplementary Information). Finally, the residual mask was removed, and the sample was immersed in a hydrofluoric acid solution to selectively etch the $SiO_2$ sacrificial layer and undercut the disk periphery.

Figure 1(d) and 1(e) show scanning electron microscope (SEM) images of a fabricated MDR with a radius of 30 μm. Both the top surface and etched sidewalls appear smooth, indicating that the optimized fabrication process significantly suppressed etching-induced roughness. A rectangular access window was also fabricated beside the disk to provide sufficient clearance for optical coupling using a dimpled tapered fiber. The tilted SEM image in Fig. 1(f) confirms that the disk periphery is successfully undercut, forming an air-suspended geometry suitable for supporting high-Q WGM resonances. Additional cross-sectional SEM characterization of a cleaved MDR before $SiO_2$ undercutting further confirmed that the etched sidewall angle was approximately 70°.

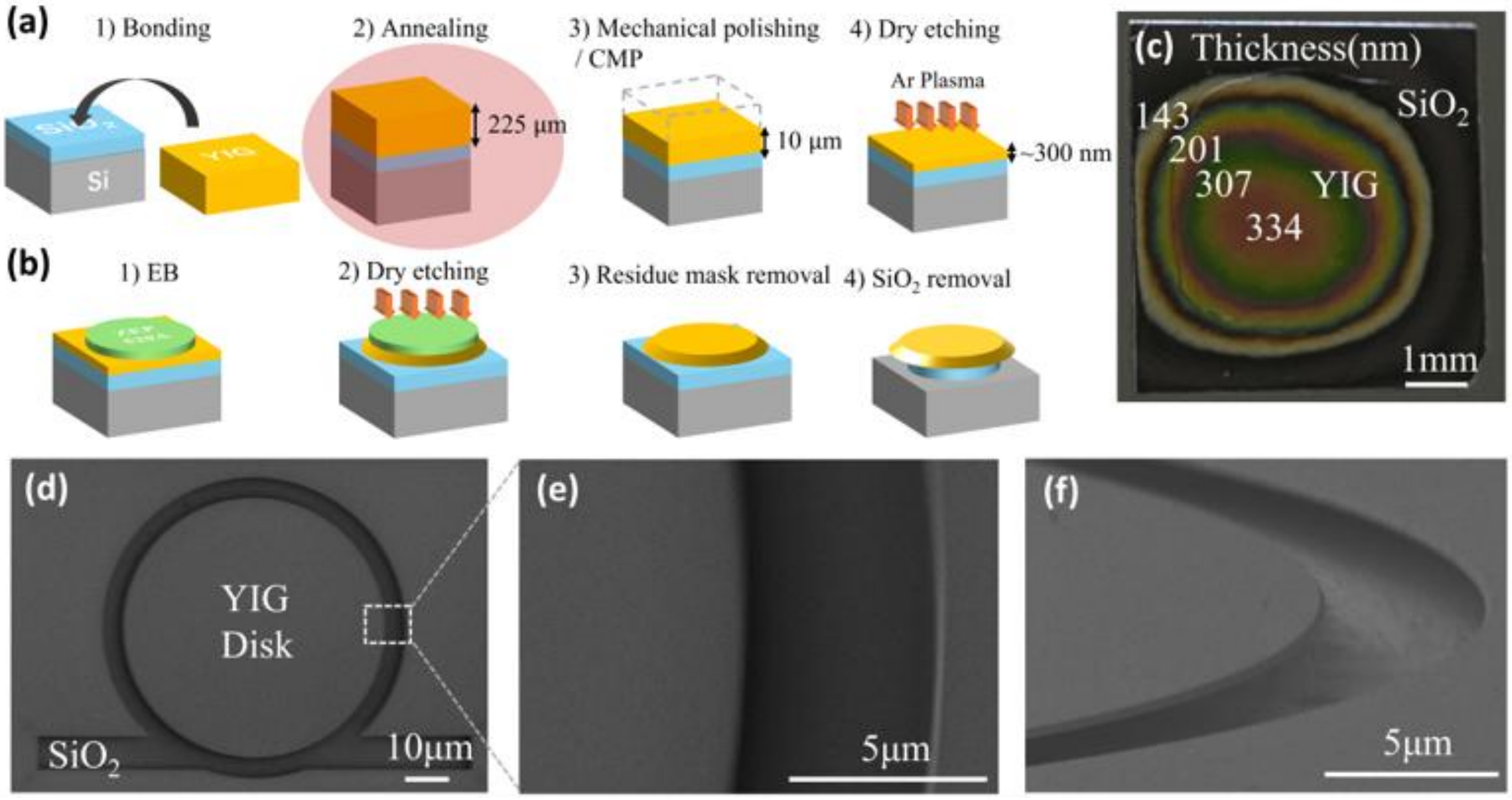

Fig. 1. Fabrication flow of (a) YIG thin films on glass and (b) a YIG-based MDR. (c) Optical microscope image of a YIG thin film on glass (YIGOI structure). (d) Scanning electron microscope (SEM) image of a fabricated MDR. (e) Magnified view of (d). (f) Tilted SEM image of the MDR.

## 3. Device Characterization

### 3.1 Transmission measurement

The fabricated Bi:YIG disks were first characterized by measuring their optical transmission spectra using a dimpled tapered fiber to couple light into and out of the MDRs. Figure 2(a) shows the measurement setup. Light from a tunable laser source (TSL-710, Santec Corporation), swept over the wavelength range of 1500–1600 nm, was sent to a fiber polarization controller. During the measurements, the polarization controller was carefully adjusted to maximize the extinction ratio of the transverse-electric(TE)-like mode resonance dips. The light was then guided into a dimpled tapered fiber, whose evanescent field overlaps with the confined WGM field of the disk, enabling optical coupling to the MDRs (Fig. 2(b)). The transmitted light collected through the same fiber was detected by a power meter (81634B, KEYSIGHT Corporation). Figure 2(c) shows the measured transmission spectrum of a fabricated Bi:YIG MDR with a thickness of approximately 320 nm. In the wavelength range of 1560–1590 nm, the spectrum exhibits periodically repeated resonance groups with a spacing of approximately 6.0 nm, corresponding to the free spectral range (FSR) between adjacent azimuthal modes within each radial mode family. Hereafter, we focus on three prominent resonance dips within each spectral period, marked by red, blue and black dots in the plot. Figure 2(d) shows a magnified spectrum of a WGM resonance at 1569.78 nm, together with a Lorentzian fit curve. The fitting yields a loaded $Q$ factor of $1.52 \times 10^5$. To the best of our knowledge, this is a record-high $Q$ factor among micro-scale YIG-based resonators and approaches the order of conventional macroscopic YIG sphere resonators. For the investigated TE-mode resonance dips, the extinction ratio reached approximately 10 dB, indicating that the MDR and the dimpled tapered fiber were operated near the critical-coupling regime. The intrinsic $Q$ factor is therefore estimated to be about a few $10^5$, which is still below the absorption-limited $Q$ factor of a few million estimated from the material absorption coefficient of 0.05 dB/cm. This discrepancy suggests that the present $Q$ factor is not limited by intrinsic material absorption, but is still governed by scattering loss originating from residual roughness on the top surface and etched sidewalls.

To identify the mode origins of the observed resonance peaks, we performed eigenmode analysis using the finite element method (FEM) with a two-dimensional axisymmetric model in COMSOL Multiphysics. The resonant frequencies, $\omega_r$, and the corresponding spatial electric-field distributions, $|\boldsymbol{E}|$, were calculated for the WGM resonances. The calculated TE radial modes shown in Fig. 2(c) provide the best explanation for the measured resonance families. In particular, the calculated FSRs, spectral ordering, and relative wavelength spacings agree well with the measured transmission spectrum. Based on this agreement, we assigned the three prominent resonance dips in each spectral period to the fundamental, second-order, and third-order radial TE-mode families. Figure 2(e) shows the simulated cross-sectional electric-field distributions of these modes. The fundamental mode is tightly confined near the outer periphery of the disk, whereas the higher-order radial modes penetrate deeper into the disk interior. Based on the FEM results, the mode volume of the fundamental TE mode was calculated to be approximately 200 $(\lambda/n)^3$. This value is about 150 times smaller than that of a conventional high-$Q$ YIG sphere resonator with a radius of 150 µm, whose mode volume is estimated to be approximately 30,000 $(\lambda/n)^3$ [22]. Combining the experimentally measured $Q$ factor with the FEM-estimated $V$ value yields a high $Q/V$ ratio of approximately 720, demonstrating an order-of-magnitude improvement in $Q/V$ for high $Q$ YIG resonators. This result highlights the advantage of the planar microdisk geometry for simultaneously achieving tight optical confinement and high $Q$ factor in monocrystalline YIG.

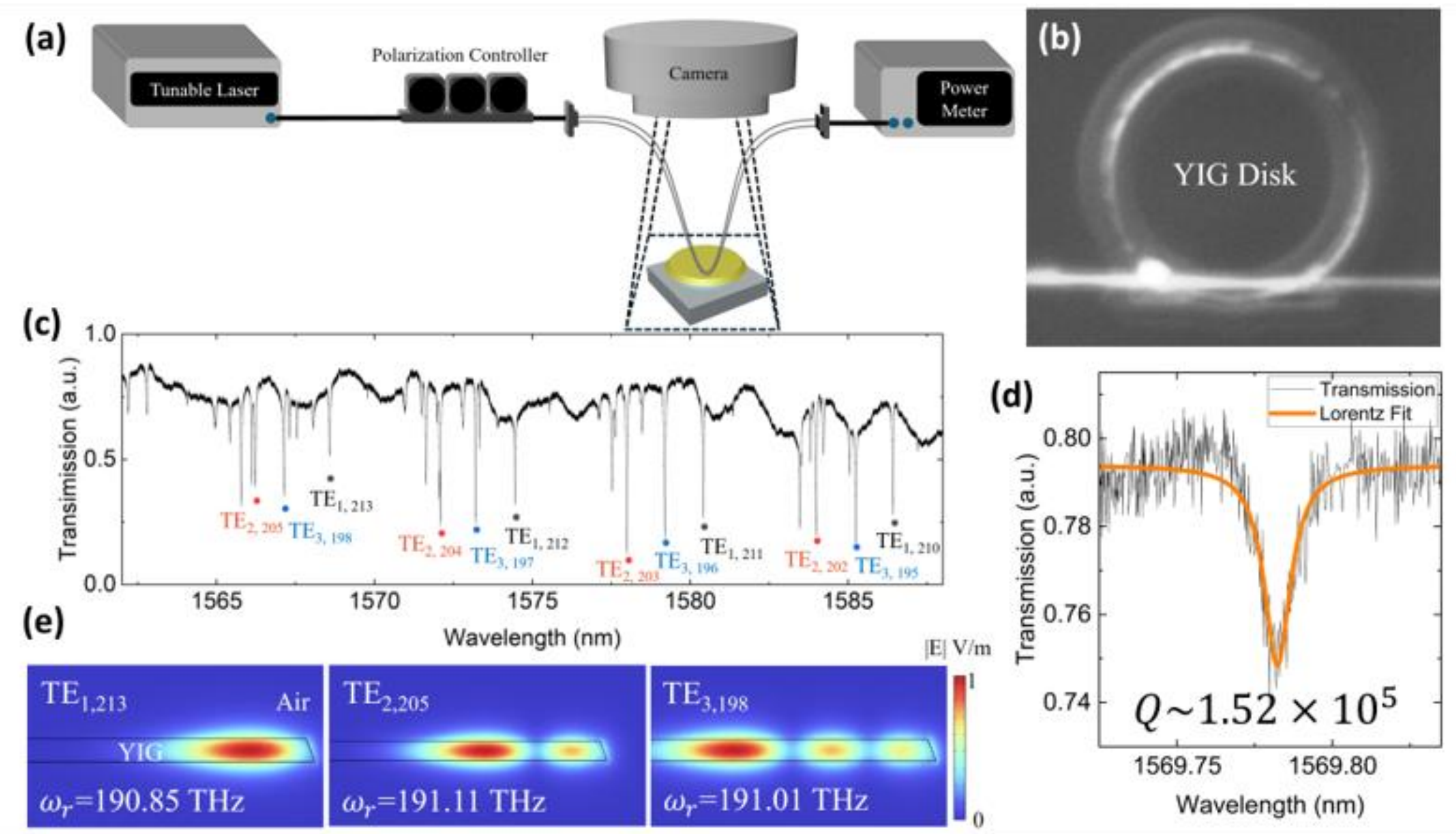

Fig. 2. (a) Schematic of the optical characterization setup. (b) Optical microscope image of the YIG disk coupled to a dimpled tapered fiber under laser illumination. (c) Measured transmission spectrum. The spectrum exhibits multiple WGM resonances. The fundamental, second-order, and third-order radial TE mode families are identified and labeled as TE_(1,m),TE_(2,m),and TE_(3,m), respectively, where m denotes the azimuthal mode number. (d) Measured resonance peak with Q~1.52×10^5. (e) Cross-sectional electric field magnitude distributions for the fundamental, second-order, and third-order radial TE modes simulated by FEM.

### 3.2 Magneto-optical effect

Next, we investigated the MO responses of the fabricated Bi:YIG MDRs under an externally applied magnetic field. The experimental setup is shown in Fig. 3(a). Bidirectional optical transmission was simultaneously probed using a dimpled fiber connected to two photodetectors (PDA Series, Thorlabs, Inc), which were connected to an oscilloscope (InfiniiVision DSOX4024A, KEYSIGHT Corporation) to monitor clockwise (CW) and counter clockwise (CCW) WGM resonances supported in the MDR. The external magnetic field was varied by translating a pair of side-by-side permanent magnets with opposite polarities beneath the MDR. Twenty-three measurement points were defined at equal intervals along the translation path. At each position, the applied magnetic-field vector was measured using a three-axis Hall-effect sensor (TLE493DW2B6MS2GOTOBO1, Infineon Technologies AG), and the magnetization inside the disk was estimated using the Stoner–Wohlfarth model[48]. Figure 3(b) and 3(c) show the measured magnetic-field components and the corresponding estimated magnetization components at each magnet position. The estimated magnetization exhibits relatively strong in-plane components and a suppressed out-of-plane component. This behavior reflects the shape anisotropy of the thin disk, which produces a large demagnetizing field along the out-of-plane direction and makes the disk plane the magnetic easy plane.

For the MO experiment, we used a Bi:YIG disk with a radius of 30 μm and a thickness of 250 nm, and selected a fundamental TE WGM resonance at a wavelength of 1571.62 nm with an azimuthal mode number of $m$ = 198. Figure 4(a) shows the CW and CCW mode spectra recorded using the oscilloscope as the magnetic field was varied. Two distinct spectral responses were observed. First, both the CW and CCW resonance frequencies exhibited a synchronized common shift depending on the magnet position. The extracted common frequency shifts are plotted in black in Fig. 4(b). Second, relative frequency shifts emerged between the CW and CCW modes. This frequency difference, $\Delta f_{\mathrm{CCW-CW}}$, is plotted in black in Fig. 4(c). The narrow linewidth of the high-$Q$ WGM resonance allowed us to resolve these magnetic-field-induced spectral responses in the MDR.

To clarify the physical origins of the two distinct spectral responses, we compared the experimental results with a theoretical model based on perturbation theory for MO resonators.

We first analyzed the common frequency shift, which appeared identically in the CW and CCW modes, suggesting a reciprocal origin. We attribute this common shift to the second order MO response, particularly the Cotton-Mouton effect, which is governed by the magnetization-dependent diagonal elements of the material permittivity tensor. In this framework, the frequency shift can be analytically described by the overlap integral between the in-plane electric field $E_{\mathrm{in}}$ and the in-plane magnetization $M_{\mathrm{in}}$:

$$\Delta\omega = \omega_0 \pi \frac{\iint r[(G_{12} - G_{11}) M_{\mathrm{in}}^2\,]|\boldsymbol{E}_{\mathrm{in}}|^2\, dr\, dz}{\iint r\, \varepsilon_r |\boldsymbol{E}|^2\, dr\, dz} \quad (1)$$

where $\omega_0$ is the unperturbed resonance frequency, $r$ and $z$ are the radial and vertical coordinates, $G_{ij}$ are the elements of the quadratic MO tensor (See Supplementary), $\varepsilon_r$ is diagonal element of the relative permittivity of Bi:YIG, and $|\boldsymbol{E}|$ is the local electric field amplitude. By substituting the magnetization estimated from Stoner-Wohlfarth model into this equation and using a fitted parameter of $(G_{12} - G_{11}) M_s^2 \sim 8.0 \times 10^{-4}$ , where $M_s$ is the saturation magnetization, we calculated the theoretical shift, which is plotted in red in Fig. 4(b). The theoretical curve well reproduces both the overall trend and the magnitude of the experimental data, confirming that Cotton-Mouton effect dominates the reciprocal common shift.

In contrast, the relative frequency shift between the CW and CCW modes indicates a non-reciprocal origin. We attribute this shift to the nonreciprocal phase shift (NRPS), which originates from the off-diagonal elements of the MO permittivity tensor. The relative frequency difference, $\Delta f_{\mathrm{CCW-CW}}$, is given by[49]:

$$\Delta f_{\mathrm{CCW-CW}} = \omega_0 \frac{\iint r\, f M_Z \left(E_r^* E_\varphi - E_\varphi^* E_r\right) dr\, dz}{\iint r\, \varepsilon_r |\boldsymbol{E}|^2\, dr\, dz} \quad (2)$$

where $M_z$ is the out-of-plane magnetization, and $E_r$and $E_\varphi$ are the radial and azimuthal components of the electric field, $f$ is related to the specific faraday rotation angle $\theta_f$ per magnetization by $f = 2\sqrt{\varepsilon_r}\theta_f/(k_0 M_s)$, where $k_0$ is the vacuum wavenumber. Here, the cross-product term $(E_r^* E_\varphi - E_\varphi^* E_r)$corresponds to the $S_3$ component of the Stokes parameters, which quantifies the difference between the right and left circularly polarized components.

The red curve in Fig. 4(c) shows the calculated $\Delta f_{\mathrm{CCW-CW}}$, which captures both the overall trend and the magnitude of the experimentally measured frequency difference, demonstrating the observed nonreciprocal frequency difference was induced by the MO NRPS effect. In this calculation, the peak faraday rotation angle was set to 2017 deg/cm to reproduce the measured frequency difference. This value is about twice the faraday rotation measured for the bulk material. Moreover, the estimated out-of-plane magnetization giving the peak faraday rotation was $M_z = 647$ Gs, which is likely smaller than the saturation magnetization because only a moderate magnetic field was applied along the hard axis of the thin MDR. These observations suggest that the effective MO response is largely enhanced in the microfabricated Bi:YIG MDR. One possible origin is surface-symmetry breaking at the air–garnet interface, which has been reported to significantly amplify the local MO response[50]. We note that a constant, reciprocal frequency differences of approximately 100 MHz was introduced to the theoretical curve. This initial shift, present even without an effective magnetic field, is attributed to the inherent CW-CCW mode coupling induced by surface scatterers or fabrication defects.

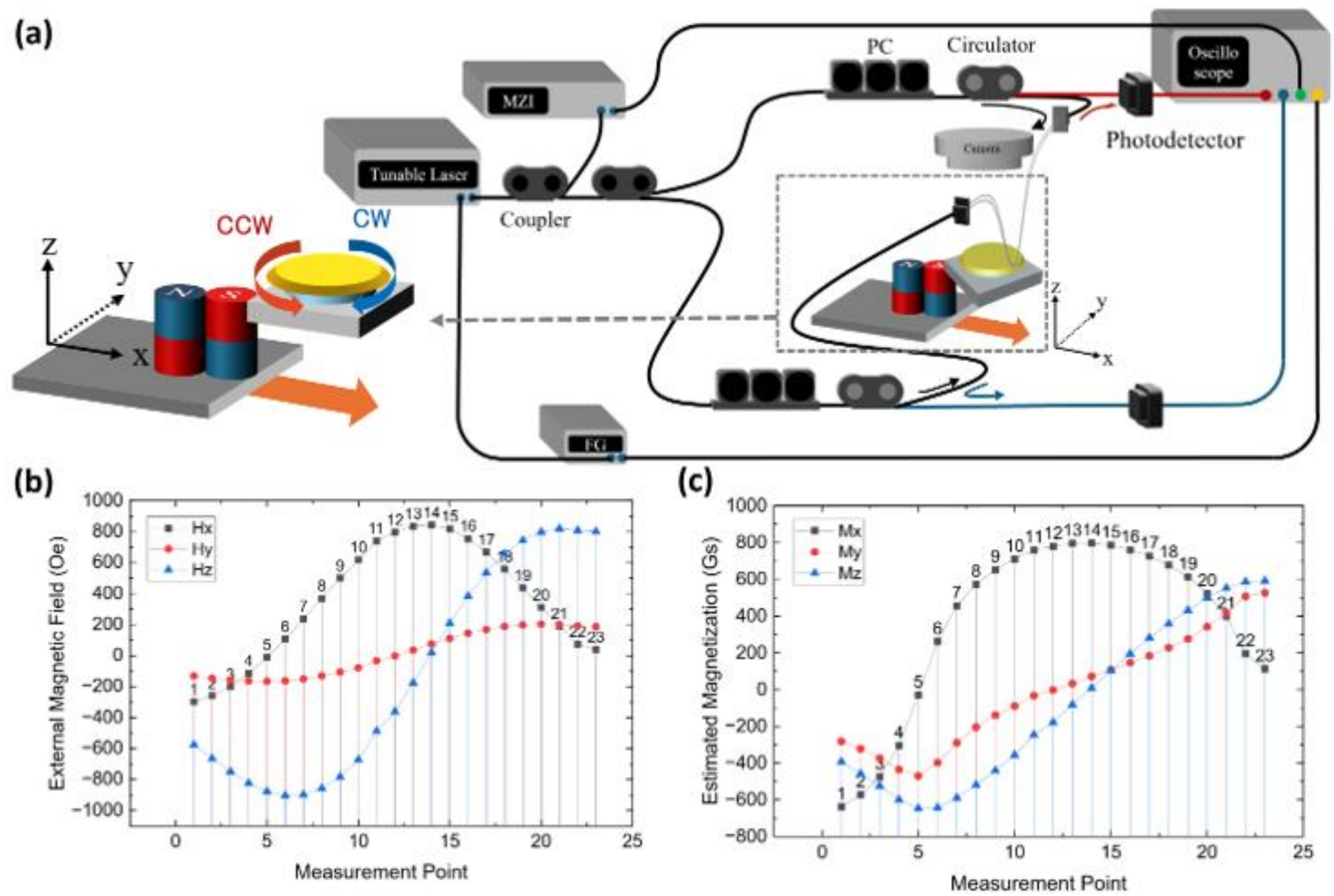

Fig. 3. (a) Schematic illustration of the experimental setup. Light from a tunable laser is coupled into the resonator via a dimpled tapered fiber to excite CW and CCW WGMs. A Mach-Zehnder interferometer (MZI) is used for wavelength calibration. An external magnetic field is applied using a pair of permanent magnets mounted on a motorized translation stage, allowing the magnetic field at the disk position to be continuously varied by moving the stage along the x-axis. (b) External magnetic field applied to the MDR when varying magnet position. (c) Estimated magnetization in the MDR at each measurement point.

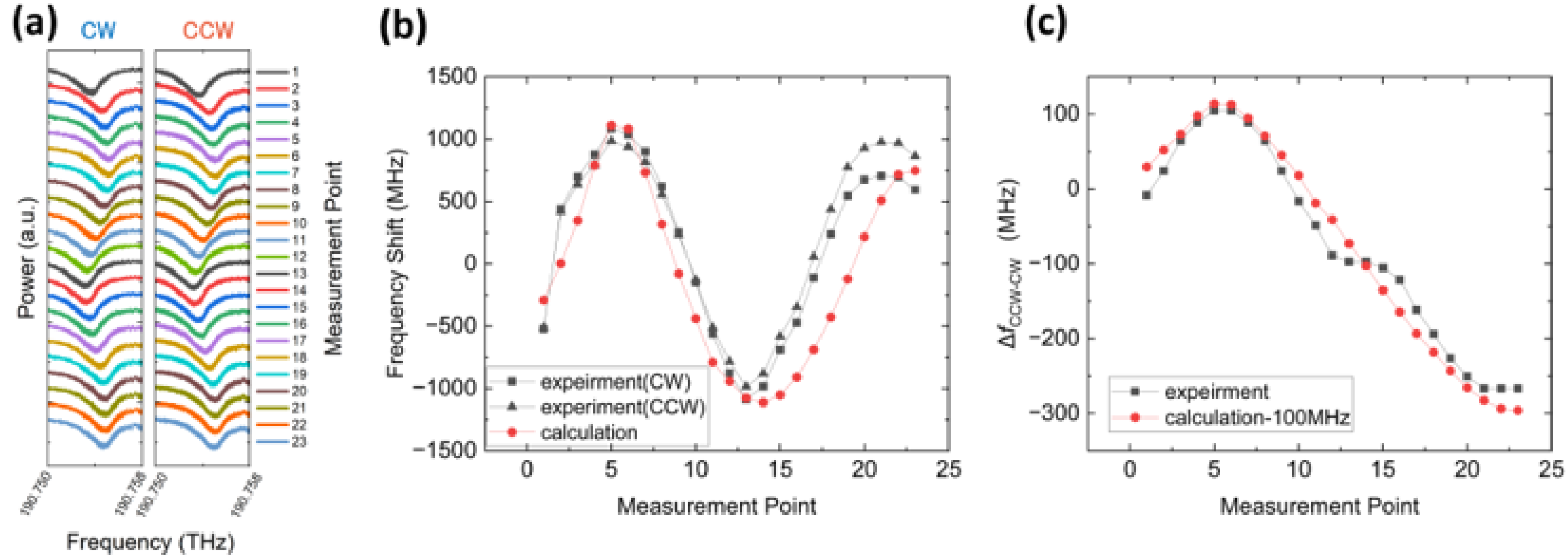

Fig. 4. (a) Transmission spectra for the CW and CCW WGM at each magnet position. (b) Resonance frequency shifts of the CW (black squares) and CCW (black triangle) modes. The frequency shifts calculated from the model are plotted as the red data points. (c) Measured non-reciprocal frequency difference, defined as the difference between the CW and CCW resonance frequencies (black squares). The red points are of the theoretical model, plotted with a 100 MHz offset.

## 4. Discussion

The demonstrated NRPS in the Bi:YIG MDRs suggests a route toward compact on-chip optical isolators. Compared with the pioneering work by Bi et al.[18], which used a racetrack resonator with a footprint of approximately 24,400 $\mu m^2$, our MDR has a footprint of approximately 2,830 $\mu m^2$ for a radius of 30 μm, corresponding to nearly an order-of-magnitude reduction in device area. Presuming the experimentally observed NRPS of ~267 MHz, a loaded $Q$ factor of approaching $10^6$ would be required for the frequency splitting to surpass the resonance linewidth, which is a prerequisite for isolator operation. This required $Q$ factor would be reduced by approximately 20% if the material were fully vertically magnetized. Such a $Q$ factor should be accessible by further suppressing scattering loss through improved fabrication processes, such as additional surface smoothing or chemical mechanical polishing.

Another promising route to reduce the required $Q$ factor is to increase the NRPS itself by engineering the transverse spin-density ($S_3$ Stokes parameter) distribution of the optical mode. The magnitude of the NRPS is governed by the spatial overlap between the $S_3$ distribution and the MO medium. In bare air-cladded MDRs, the fundamental TE mode at 1.55 μm exhibits two pronounced $S_3$ peaks with opposite signs inside the disk (Fig. 5(a, b)). These opposite contributions largely cancel each other after spatial integration, yielding a relatively weak net NRPS. To mitigate this unwanted cancellation, we propose manipulating the $S_3$ distribution by introducing a non-magnetic cladding layer, which could be deposited using techniques such as atomic layer deposition (ALD). Figures 5(c, d) show the optical field and $S_3$ distributions of a cladded MDR with a radius of 30 μm and a Bi:YIG thickness of 250 nm. The addition of a non-magnetic 600-nm-thick clad layer with a refractive index of 2.2 pulls the optical mode outward, together with the $S_3$ distribution. In the modified MDR, only one of the two $S_3$ peaks retains a large overlap with the MO material, thereby suppressing the cancellation between opposite-sign contributions to the NRPS. Figure 5(e) summarizes calculated NRPS-induced resonance splitting for the fundamental TE mode as a function of the cladding thickness $w$ for three different cladding refractive indices (1.8, 2.2, and 2.6). These results show that both $w$ and the refractive index of the cladding material need to be controlled to enhance the nonreciprocal frequency splitting. At the optimal value of $w$ = 600 nm for the clad refractive index of 2.2, the NRPS-induced splitting reaches the maximum value of ~ 5.4 GHz, which is ~ 20 times larger than that of the air-cladded MDR corresponding to $w = 0$ and far exceed the measured linewidth (~1.25 GHz) of the high $Q$ WGM resonance in our device. This massive MO response provides a route for magnetic-field-controlled dispersion engineering[51], paving the way for novel microcombs whose modal dispersion, including local anomalous dispersion induced by polarization-mode coupling, can be tuned after fabrication using an external magnetic field.

High-$Q/V$ YIG-based MDRs are also promising for microwave-to-optical conversion mediated by magnons, owing to the tight spatial confinement of both photons and magnons. Recent theoretical studies have predicted that thin YIG disk and ring microcavities can drastically improve the spatial overlap between optical WGMs and magnon Kittle modes, increasing the magnon-photon coupling rate into the kilohertz range[52]. This value is two orders of magnitude larger than the coupling rates typically achieved in conventional YIG spheres. The demonstrated monocrystalline YIG platform is particularly favorable for such applications, because its high crystalline quality should help preserve the low magnetic damping and long spin coherence.

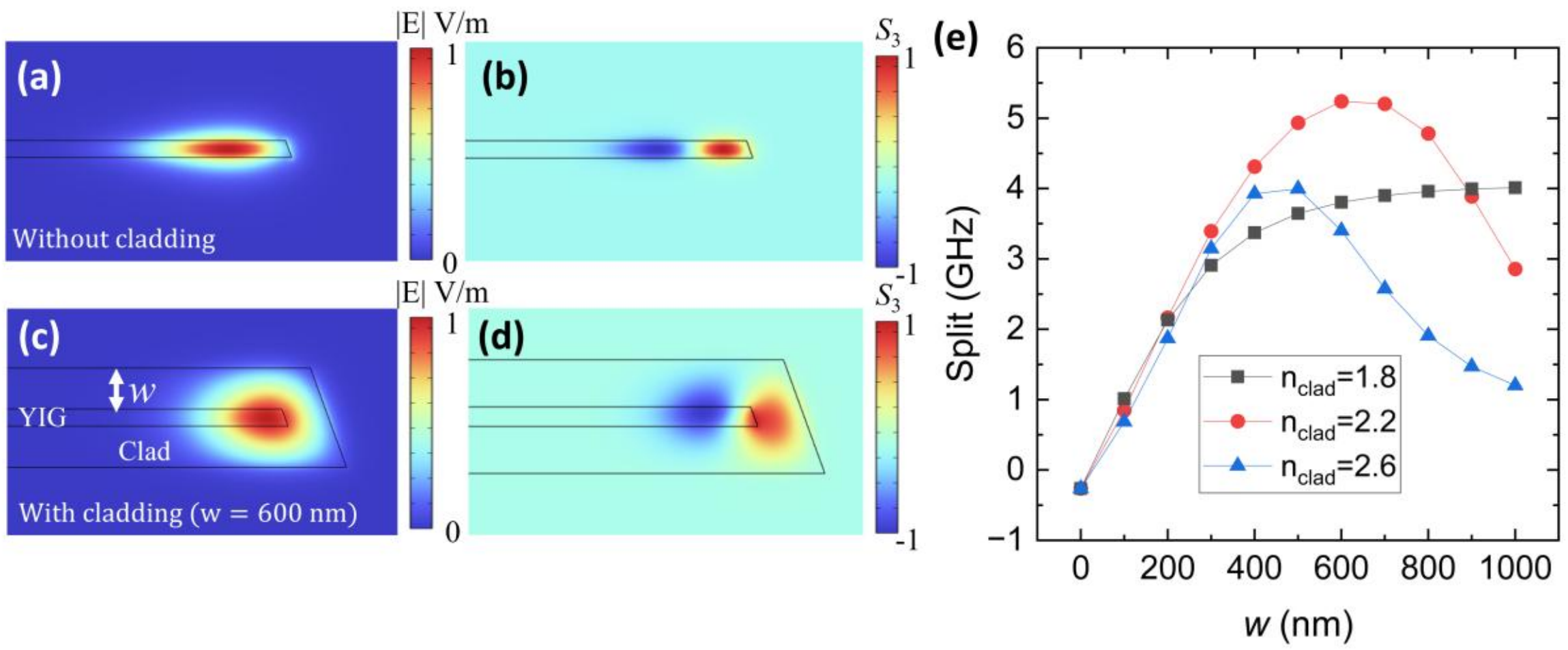

Fig. 5. Simulated electric-field intensity and transverse spin-density ($S_3$) distributions for the fundamental TE mode (a, b) under air cladding ($w = 0$ nm), and (c, d) with a non-magnetic cladding layer ($w = 600$ nm, $n_{clad} = 2.2$). The presence of the dielectric cladding pulls the optical mode outward, shifting one of the oppositely signed $S_3$ components into the non-MO cladding region. (e) Calculated NRPS-induced resonance splitting as a function of cladding thickness $w$ for three different cladding refractive indices ($n_{clad} = 1.8, 2.2,$ and $2.6$).

## 5. Conlclusion

In conclusion, we have demonstrated high-$Q$ MO MDRs based on monocrystalline Bi:YIG. We developed a fabrication process for Bi:YIG-based MDRs and observed a WGM resonance with a high $Q$ factor of approximately $1.52 \times 10^5$ within a small $V$ of $200\ (\lambda/n)^3$, corresponding to a high $Q/V$ ratio of 760. We also experimentally investigated the MO response of the MDR under an applied magnetic field. Using analytical models, we attributed the observed reciprocal resonance shifts to the Cotton–Mouton effect and the frequency splitting between the CW and CCW modes to NRPS. Our analysis further indicates that the effective MO response of the microstructured Bi:YIG can be significantly enhanced relative to that expected from the bulk material parameters. We also discussed dielectric-cladded MDRs for enhancing NRPS effect in WGM resonances. These findings highlight monocrystalline YIG MDRs as a promising MO platform offering high $Q/V$ ratios and represent an important step toward their use in a broad range of applications, including compact integrated optical isolators, magnetically tunable microcombs, and efficient bidirectional microwave-to-optical conversion mediated by magnons.

**Funding.** This work was supported by JST FOREST (JPMJFR213F), JST CREST (JPMJCR19T1) and KAKENHI(25K01697, 24K17582), Iketani Foundation, and Nippon Sheet Glass Foundation.

**Acknowledgment.** We would like to thank H. Otsuki, H. Matsukiyo, M. Nishioka, S. Ishida, R. Hisatomi, K. Usami K. Yamaguchi, H. Moriguchi, and A. I. Musorin for their technical support and fruitful discussions.